\documentclass[aps,prl,twocolumn,showpacs]{revtex4-1}
\usepackage{graphicx}
\begin{document}

\title{Enhancement of the Superconducting Transition Temperature in FeSe Epitaxial Thin Films by Anisotropic Compression}

\author{
Fuyuki Nabeshima$^{1}$, Yoshinori Imai$^{1}$, Masafumi Hanawa$^{2}$, Ichiro Tsukada$^{2}$, and Atsutaka Maeda$^{1}$
}
\affiliation{
$^{1}$Department of Basic Science, the University of Tokyo, Meguro, Tokyo 153-8902, Japan\\
$^{2}$Central Research Institute of Electric Power Industry, Yokosuka, Kanagawa 240-0196, Japan\\
}

\date{\today}

\begin{abstract}

In order to investigate the effects of in-plane strain on the superconductivity of FeSe, epitaxial thin films of FeSe were fabricated on CaF$_2$ substrates. The films are compressed along the $a$-axis and their superconducting transition temperatures $T_{\mathrm c}^{\mathrm {zero}}$ reach 11.4 K, which is approximately 1.5 times higher than that of bulk crystals. The $T_{\mathrm c}$ values are weakly dependent on the ratio of the lattice constants, $c$/$a$, compared to that of Fe(Se,Te). Our results indicate that even a binary system FeSe has room for improvement, and will open a new route for the application of Fe-based superconductors.

\end{abstract}


\maketitle

The discovery of iron-based materials with a high superconducting transition temperature $T_{\mathrm {c}}$ has attracted much attention for both fundamental studies and practical applications.~\cite{Kamihara}
Iron-chalcogenide superconductors~\cite{HsuFeSe} (Fe$Ch$), Fe(Se,Te), have the simplest crystal structure, consisting of only two-dimensional conducting planes. 
Although the  $T_{\mathrm {c}}$ values of these materials are low compared with those of other families, the $T_{\mathrm {c}}$ values are strongly dependent on the applied pressure.
In fact, the onset of the resistive transition, $T_{\mathrm {c}}^{\mathrm {onset}}$, reaches 37 K under a hydrostatic pressure of approximately 4 GPa.~\cite{JPSJ.78.063704,nmat2491}
Therefore, anisotropic pressure effects on $T_{\mathrm {c}}$ of Fe$Ch$ are of great interest.

Studies on the film growth of the optimally doped Fe(Se,Te) have suggested that the in-plane ($ab$-plane) compressive strain can increase the $T_{\mathrm {c}}$ value above that of bulk crystals; such studies have also found enhanced superconducting properties.~\cite{bellingeri:102512,iida:202503,APEX.4.053101,WSiNatCom} 
Additionally, under a hydrostatic pressure,~\cite{JPSJ.78.063705} FeSe shows a large increase in $T_{\mathrm {c}}$ compared to optimally doped Fe(Se,Te). 
Thus, we expect FeSe thin films to have $T_{\mathrm {c}}$ values higher than those of bulk single crystals when in-plane compressive strain is successfully introduced in the films.
Recently, high $T_{\mathrm {c}}$ superconductivity was reported in single-unit-cell-thick FeSe films on SrTiO$_3$ substrates.~\cite{0256-307X-29-3-037402} 
Although it is unclear whether this phenomenon is characteristic of the interface, this report provides another example showing that FeSe has potential as a very high $T_{\mathrm {c}}$ superconductor.
Thus, a very important question is whether we can realize this high $T_{\mathrm {c}}$ superconductivity as a bulk nature.
Additionally, from this viewpoint, we should investigate the effects of anisotropic strain in FeSe films.

Several groups have reported the growth of FeSe films using oxide substrates.~\cite{0953-8984-21-23-235702,nie:242505,PhysRevLett.103.117002,jourdan:023913,Jung20101977,Chen2011515,PhysRevLett.108.257003} 
However, the $T_{\mathrm {c}}$ values of the FeSe films reported to date are rather low, and there are few reports on the fabrication of FeSe thin films with good superconducting properties. 
The lattice constants of these FeSe films were similar to those of bulk crystals. 
In a previous study of Fe(Se,Te) thin film fabrication,~\cite{APEX.3.043102} we found that $T_{\mathrm {c}}$ is positively correlated with the ratio of the lattice parameters, $c/a$. 
A subsequent study~\cite{APEX.4.053101} revealed that compared with oxide substrates, the use of CaF$_2$ substrates can introduce a strong in-plane compressive strain in the films. 
We expect to observe the same effect for FeSe films and also expect an enhancement of superconducting properties, such as an increased $T_{\mathrm {c}}$.

In this Letter, we report the fabrication of high-quality epitaxial thin films of FeSe on CaF$_2$ substrates using a pulsed laser deposition (PLD) method. 
We demonstrate that the films are compressed along the $a$-axis and their superconducting transition temperatures $T_{\mathrm c}^{\mathrm {zero}}$ reach 11.4 K, which is approximately 1.5 times higher than that of bulk crystals.  
Our results in this binary system are very promising, and will open a new route for the application of Fe-based superconductors.


All of the films in this study were grown by the PLD method with a KrF laser.~\cite{JJAP.49.023101,APEX.3.043102} 
FeSe polycrystalline pellets were used as targets. 
The substrate temperature, the laser repetition rate, and the back pressure were 280$^\circ \mathrm{C}$, 10 Hz and 10$^{-6}$\ Torr, respectively. 
Commercially available CaF$_2$ (100) substrates were used for the present experiments. 
Although some groups have reported that crystal orientation along the (101) direction, accomplished by using substrate temperatures of as high as 500$^\circ$C, is the key for the fabrication of FeSe thin films with high $T_{\mathrm {c}}$ values,~\cite{PhysRevLett.103.117002,Jung20101977} we adopted lower substrate temperatures and obtained $c$-axis preferred orientation. 
Nevertheless, as will be described later, our films show very good superconducting properties.
Indeed, films with $c$-axis orientation are advantageous for measurements of the in-plane ($ab$-plane) conductivity.
We prepared eight thin films with different thicknesses. 
The films were fabricated in a six-terminal shape through the use of a metal mask. 
The measured area was 0.95 mm long and 0.2 mm wide.
The thicknesses of the grown films were measured using a Dektak 6M stylus profiler and were estimated to be 60 - 235\ nm.
The films are designated as C1 - C8 in order of the film thickness. 
The specifications of all of the films are summarized in table~\ref{tab:FilmSpec}.
The crystal structures and the orientations of the films were characterized by four-circle X-ray diffraction (XRD) with Cu K$\alpha$ radiation at room temperature. 
The $c$-axis and $a$-axis lattice constants of the films were calculated from the positions of the 001-004 reflections and the 200 reflection, respectively. 
The electrical resistivity was measured using a physical property measurement system (PPMS) from 2 to 300\ K under magnetic fields of up to 9\ T applied perpendicularly to the film surface. 
Superconductivity was also confirmed by the magnetization measurement.

\begin{table}[h]
	\caption{\label{tab:FilmSpec}Specifications of the grown films.}
	\begin{ruledtabular}
		\begin{tabular}{ccccccc}
		 & \shortstack{thickness \\ (nm)} & \shortstack{$a$-axis \\ (\AA)} & \shortstack{$c$-axis \\ (\AA)} & \shortstack{${T_{\mathrm c}}^{\mathrm {onset}}$ \\ (K)} & \shortstack{${T_{\mathrm c}}^{\mathrm {mid}}$ \\ (K)} & \shortstack{${T_{\mathrm c}}^{\mathrm {zero}}$ \\ (K)} \\ \hline
		C1 & 60 & 3.761 & 5.537 &  6.43& 5.42 & $<$ 4.2  \\
		C2 & 75 & 3.747 & 5.549 & 8.60 & 7.48 & 6.11 \\
		C3 & 85 & 3.720 & 5.560 & 11.67 & 10.51 & 8.41 \\
		C4 & 92 & 3.730 & 5.567 & 11.53 & 10.80 & 9.33 \\
	 	C5 & 120 & 3.715 & 5.578 & 11.71 & 11.46 & 10.82 \\
		C6 & 150 & 3.714 & 5.584 & 12.35 & 11.92 & 11.38 \\
		C7 & 205 & 3.722 & 5.582 & 11.74 & 11.44 & 10.38 \\
		C8 & 235 & 3.730 & 5.579 & 11.24 & 11.03 & 10.47 \\
		\end{tabular}
	 \end{ruledtabular}
\end{table}

\begin{figure}[ht]
\includegraphics[bb=32 46 761 558,width=25em,clip]{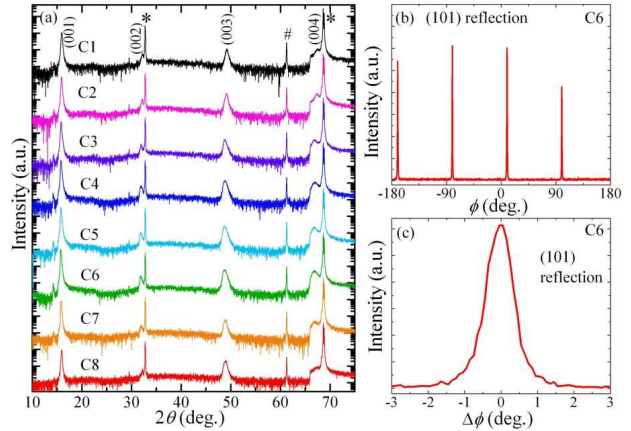}
\caption{(a) XRD patterns of $\omega$-$2\theta$ scans perpendicular to the substrate plane for the eight FeSe thin films on CaF$_2$ (C1 - C8). The asterisks represent the peaks resulting from the substrate, and the number signs represent the unidentified peaks. (b) XRD pattern of the $\phi$ scan of the 101 reflection from film C6. (c) Enlargement of (b) around the peak at approximately $\phi$ = 10 deg.}
\label{XRD}
\end{figure}

Fig. ~\ref{XRD}(a) shows the $\omega$-2$\theta$ X-ray diffraction patterns of the eight FeSe thin films grown on CaF$_2$ (100) substrates (C1 - C8). 
These films show only the 00$l$ reflections of the tetragonal PbO structure, indicating that the films are well-oriented along the $c$-axis. 
Fig. ~\ref{XRD}(b) shows the $\phi$ scans of the 101 reflection of film C6. 
A clear four-fold symmetry reflection was obtained. 
The full widths at half maximum (FWHM) is $\Delta \phi \sim 1.0^\circ$ (shown in the inset of fig. ~\ref{XRD}(b)), which is comparable to those of Fe(Se,Te) films on CaF$_2$ substrates.~\cite{APEX.4.053101} 
We have also confirmed the in-plane orientation to be FeSe[100] $\parallel$ CaF$_2$[110], similar to Fe(Se,Te).
The calculated $a$- and $c$-axis lengths of the grown films are shown in table~\ref{tab:FilmSpec}. 
It is clear that the $a$-axis lengths are dependent on the film thickness; the $a$-axis decreases as the film thickness increases up to 150 nm, and then the $a$-axis slightly increases for films with larger thickness. 
This behavior is similar to that observed in Fe(Se,Te) films on LaAlO$_3$.~\cite{bellingeri:102512}
The authors in ref. ~\cite{bellingeri:102512} explained that this thickness dependence is due to Volmer-Weber type growth of the Fe(Se,Te) film, which we believe is also the case with FeSe films on CaF$_2$. 
Compared with the lattice parameters for bulk single crystals ($a = 3.775 \pm 0.02$ \AA, $c = 5.52 \pm 0.03$ \AA),~\cite{PhysRevB.79.014522,Souza,PhysRevB.83.224502} the films grown on CaF$_2$ are compressed along the $a$-axis and are simultaneously elongated along the $c$-axis. 
Such short $a$-axis lengths of the FeSe films can not be explained simply by the difference between the lattice constants of the substrate and the overlayer because the lattice constants of CaF$_2$ ($a / \sqrt 2 = 3.863$ \AA) are longer than the $a$-axis of FeSe.
The penetration of F$^{-}$ ions from the CaF$_2$ substrates into the films has been proposed as a possible mechanism for the contraction of the $a$-axis lengths of the Fe(Se,Te) films.~\cite{IchinoseCaF2} 
The substitution of small F$^{-}$ ions for large Se$^{2-}$ ions shortens the $a$-axis.
Thus it is natural to consider that the short $a$-axis lengths of the FeSe films on CaF$_2$ can be explained by the same mechanism proposed for Fe(Se,Te) films.
We should note that the diffusion of other atoms (Fe, Te and Ca) was not detected in the previous measurement.~\cite{IchinoseCaF2}

\begin{figure}
\includegraphics[bb=17 17 324 188,width=25em,clip]{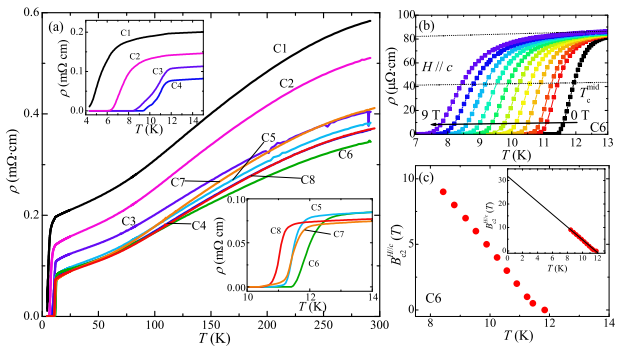}
\caption{(a) Temperature dependences of the resistivities of eight FeSe films on CaF$_2$ (C1 - C8). (b) Measurement of $\rho$ as a function of $T$ for film C6 under magnetic fields up to $\mu _0 H$ = 9\ T applied along the $c$-axis. (c) Plot of $B_{\mathrm {c2}}^{H \parallel c}$ as a function of $T_{\mathrm c} ^{\mathrm {mid}}$ for film C6. The inset shows the linear extrapolation to $T$ = 0 K.}
\label{RhoT}
\end{figure}

Fig. ~\ref{RhoT}(a) shows the temperature dependence of the resistivities of the grown films (C1 - C8). 
The temperature dependence of the resistivity shows metallic behavior similar to that of the bulk samples.
The magnitudes of the resistivity at room temperature are 0.35 - 0.6 m$\Omega \cdot$cm, which are smaller than those of bulk single crystalline samples.~\cite{PhysRevB.79.014522,Souza,PhysRevB.83.224502}
All of the films show the onset of the superconducting transition at temperatures above 4.2 K, and zero resistivity was observed in all films except film C1.
As the thickness becomes smaller for films with thickness less than 100 nm, the magnitude of the resistivity increases, and $T_{\mathrm {c}}$ decreases correspondingly.
Although the tendency that $T_{\mathrm {c}}$ decreases with decreasing thickness has also been reported for FeSe films with $c$-axis preferred orientation~\cite{PhysRevLett.108.257003} before, our films on CaF$_2$ show higher $T_{\mathrm {c}}$ even for films with much smaller thickness than those reported in ref. ~\cite{PhysRevLett.108.257003}. 
It should be noted that for FeSe films with thicknesses less than 100 nm, $T_{\mathrm {c}}$ = 9.33 K is also far better than the previously reported results for films on oxide substrates with (101) orientation ($T_{\mathrm {c}}^{\mathrm {zero}} \sim$ 6.5 K for 140 nm, $T_{\mathrm {c}}^{\mathrm {zero}} \sim$ 8 K for 1.5 $\mu$m).~\cite{PhysRevLett.103.117002,Jung20101977}
The thickness dependence of $T_{\mathrm {c}}$ is similar to that of the $a$-axis length of the films; $T_{\mathrm {c}}$ increases with increasing film thickness up to 150 nm and further increase of thickness results in the decrease in $T_{\mathrm {c}}$.
It is remarkable that films C4 - C8 show $T_{\mathrm {c}}^{\mathrm {zero}}$ values higher than those of the bulk crystals. 
In particular, film C6 has a $T_{\mathrm {c}}^{\mathrm {zero}}$ of 11.4 K, which is  approximately 1.5 times higher than those of the bulk samples. 
This is the first report demonstrating that FeSe films show higher $T_{\mathrm {c}}$ values than bulk single crystals, except for possible interface superconductivity~\cite{0256-307X-29-3-037402} between FeSe and SrTiO$_3$.
It should also be noted that these films do not require any buffer layers.

Fig. ~\ref{RhoT}(b) shows the temperature dependence of the resistivity of film C6 under magnetic fields up to $\mu _0 H= 9$\ T applied along the $c$-axis. 
The superconducting transition temperature decreases as the magnetic field increases. 
In fig. \ref{RhoT}(c) the upper critical field, $B_{\mathrm {c2}}^{H \parallel c}$, is plotted as a function of $T_{\mathrm {c}}^{\mathrm {mid}}$, the temperature at which the resistivity drops to a half of its value in the normal state. 
$B_{\mathrm c2}^{H \parallel c}$ increases almost linearly as the temperature decreases. 
The slight positive curvature of $B_{\mathrm c2}$($T$) observed at very low fields may derive from the multiband nature of this material.~\cite{Bi2Pd} 
For multiband superconductors, it is hard to predict the upper critical field at 0\ K, $B_{\mathrm c2,0}^{H \parallel c}$, from the low field data. 
Nevertheless, we show some estimated values in order to compare them to the data in the literature.
The upper critical field at 0\ K is estimated by linear extrapolation to be $B_{\mathrm c2,0}^{H \parallel c}= 33$\ T.
The conventional Werthamer-Helfand-Hohenberg theory ($B_{\mathrm c2,0} = - 0.69T_{\mathrm c} ({\mathrm d} B_{\mathrm c2} / {\mathrm d} T) \bigr| _{T = T_{\mathrm c}} $) predicts that $B_{\mathrm c2,0}^{H \parallel c} =$ 22.8\ T; this value is higher than any reported value for FeSe thin films that was estimated in the same way.~\cite{jourdan:023913,Chen2011515}  
Using $B_{\mathrm c2,0}^{H \parallel c} = \mathit{\Phi}_0 / 2 \pi \xi _{ab} (0) ^2$, we can get a Ginzburg-Landau coherence length of $\xi_{\mathrm {ab}}(0) \sim 38.0$ \AA, which is longer than that of FeSe$_{0.5}$Te$_{0.5}$ thin films on CaF$_2$ substrates.~\cite{APEX.4.053101}
These aspects of FeSe thin films on CaF$_2$ should be beneficial for applications of these materials.
Our results for this binary system will open a new path for applied studies of iron-based superconductors.

\begin{figure}
\includegraphics[bb=0 0 820 640,width=23em,clip]{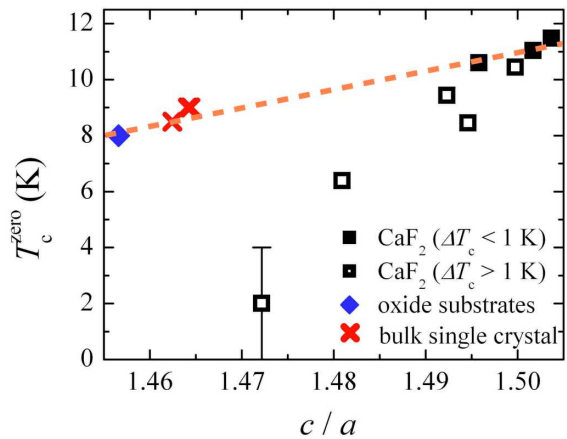}
\caption{${T_{\mathrm c}}^{\mathrm {zero}}$ values for eight FeSe films on CaF$_2$ (C1 - C8) as a function of $c/a$. The values for bulk single crystals~\cite{PhysRevB.79.014522,Souza,PhysRevB.83.224502} and films on oxide substrates~\cite{Jung20101977} are also plotted for comparison.}
\label{VScovera}
\end{figure}

Finally we discuss the relation between $T_{\mathrm {c}}$ and the lattice parameters.
As previously described, we expected that the in-plane compression of the films, represented by large $c/a$ values, would increase $T_{\mathrm {c}}$.
Indeed our results support this expectation.
In fig.~\ref{VScovera}, we plot the $c/a$ dependence of $T_{\mathrm {c}}^{\mathrm {zero}}$ for the eight films and for bulk single crystals~\cite{PhysRevB.79.014522,Souza,PhysRevB.83.224502} and films on oxide substrates.~\cite{Jung20101977}
$T_{\mathrm {c}}^{\mathrm {zero}}$ of the FeSe films on CaF$_2$ increases monotonically as $c/a$ increases, similar to that observed for FeSe$_{0.5}$Te$_{0.5}$ films.~\cite{APEX.3.043102}
However, in contrast to FeSe$_{0.5}$Te$_{0.5}$, the $T_{\mathrm {c}}$ values of single crystals and films on oxide substrates do not seem to exhibit the same dependence on $c/a$.
One possible explanation for this new complexity is the effect of disorders.
It is well-known that the excess Fe, which occupies an additional Fe site, strongly affects the superconductivity in FeSe.~\cite{PhysRevB.79.014522} 
We speculate that because of the high vapor pressure of Se, the ratio of Se to Fe becomes smaller than the stoichiometric value near the surface of the film, leading to the increase in the amount of Fe which occupies the extra Fe site.
Our result is that resistivity increases and $T_{\mathrm {c}}$ decreases with decreasing thickness for films with thickness smaller than 100 nm. This is interpreted that the influence of the regions with excess Fe becomes dominant as the films gets thinner. 
If we consider the superconducting transition width, $\Delta T_{\mathrm {c}}$ ($ = T_{\mathrm {c}}^{\mathrm {onset}} - T_{\mathrm {c}}^{\mathrm {zero}}$), to be an index of the crystal quality and consider only the data for films with $\Delta T_{\mathrm {c}} < 1$ K, $T_{\mathrm {c}}$ seems to correlate with $c/a$, as shown by an orange dashed line in fig. ~\ref{VScovera}.
In this case, $c/a$ dependence of $T_{\mathrm {c}}$ of FeSe is weak compared to that of Fe(Se,Te), and the $T_{\mathrm {c}}$ values of FeSe films under anisotropic compression are unlikely to reach 37 K, which is accomplished by hydrostatic pressure.~\cite{JPSJ.78.063704,nmat2491}
Therefore, the simultaneous compression of the $c$- and $a$-axis lengths may be a key to realize a very high $T_{\mathrm {c}}$ material.
We should note, however, that the relation between $T_{\mathrm {c}}$ and $c/a$ is simplistic and is only a guiding principle in the search for high $T_{\mathrm {c}}$ materials.
To understand the relation between $T_{\mathrm {c}}$ and crystal structure, we should evaluate specific structural parameters, such as the $Ch$-Fe-$Ch$ angle~\cite{JPSJ.77.083704} and/or the $Ch$ height from the Fe layer.~\cite{JPSJ.79.102001}

In summary, in order to investigate the effects of in-plane strain on the superconductivity of FeSe, we fabricated high-quality FeSe epitaxial thin films oriented along the $c$-axis on CaF$_2$ substrates. 
X-ray diffraction analysis showed that our films have shorter $a$-axis and longer $c$-axis lengths in comparison to bulk single crystals, demonstrating that a large in-plane compressive strain was introduced in the FeSe films. 
We demonstrated that $T_{\mathrm c}^{\mathrm {zero}}$ can reach 11.4 K, which is approximately 1.5 times greater than the $T_{\mathrm c}^{\mathrm {zero}}$ values of bulk crystals.
Further studies of strain effects will lead to higher $T_{\mathrm {c}}$ values.
Our results in this binary system are very promising and will open a new route for the application of Fe-based superconductors.


\begin{acknowledgments}
We would like to thank S. Komiya and A. Ichinose for fruitful discussion. 
We also thank K. Fukawa at the Institute of Engineering Innovation, School of Engineering, University of Tokyo for supporting in the XRD measurements of the films.
This research was supported by the Strategic International Collaboration Research Program (SICORP), Japan Science and Technology Agency.
\end{acknowledgments}


\begin{thebibliography}{26}%
\makeatletter
\providecommand \@ifxundefined [1]{%
 \@ifx{#1\undefined}
}%
\providecommand \@ifnum [1]{%
 \ifnum #1\expandafter \@firstoftwo
 \else \expandafter \@secondoftwo
 \fi
}%
\providecommand \@ifx [1]{%
 \ifx #1\expandafter \@firstoftwo
 \else \expandafter \@secondoftwo
 \fi
}%
\providecommand \natexlab [1]{#1}%
\providecommand \enquote  [1]{``#1''}%
\providecommand \bibnamefont  [1]{#1}%
\providecommand \bibfnamefont [1]{#1}%
\providecommand \citenamefont [1]{#1}%
\providecommand \href@noop [0]{\@secondoftwo}%
\providecommand \href [0]{\begingroup \@sanitize@url \@href}%
\providecommand \@href[1]{\@@startlink{#1}\@@href}%
\providecommand \@@href[1]{\endgroup#1\@@endlink}%
\providecommand \@sanitize@url [0]{\catcode `\\12\catcode `\$12\catcode
  `\&12\catcode `\#12\catcode `\^12\catcode `\_12\catcode `\%12\relax}%
\providecommand \@@startlink[1]{}%
\providecommand \@@endlink[0]{}%
\providecommand \url  [0]{\begingroup\@sanitize@url \@url }%
\providecommand \@url [1]{\endgroup\@href {#1}{\urlprefix }}%
\providecommand \urlprefix  [0]{URL }%
\providecommand \Eprint [0]{\href }%
\providecommand \doibase [0]{http://dx.doi.org/}%
\providecommand \selectlanguage [0]{\@gobble}%
\providecommand \bibinfo  [0]{\@secondoftwo}%
\providecommand \bibfield  [0]{\@secondoftwo}%
\providecommand \translation [1]{[#1]}%
\providecommand \BibitemOpen [0]{}%
\providecommand \bibitemStop [0]{}%
\providecommand \bibitemNoStop [0]{.\EOS\space}%
\providecommand \EOS [0]{\spacefactor3000\relax}%
\providecommand \BibitemShut  [1]{\csname bibitem#1\endcsname}%
\let\auto@bib@innerbib\@empty
\bibitem [{\citenamefont {Kamihara}\ \emph {et~al.}(2008)\citenamefont
  {Kamihara}, \citenamefont {Watanabe}, \citenamefont {Hirano},\ and\
  \citenamefont {Hosono}}]{Kamihara}%
  \BibitemOpen
  \bibfield  {author} {\bibinfo {author} {\bibfnamefont {Y.}~\bibnamefont
  {Kamihara}}, \bibinfo {author} {\bibfnamefont {T.}~\bibnamefont {Watanabe}},
  \bibinfo {author} {\bibfnamefont {M.}~\bibnamefont {Hirano}}, \ and\ \bibinfo
  {author} {\bibfnamefont {H.}~\bibnamefont {Hosono}},\ }\href {\doibase
  10.1021/ja800073m} {\bibfield  {journal} {\bibinfo  {journal} {Journal of the
  American Chemical Society}\ }\textbf {\bibinfo {volume} {130}},\ \bibinfo
  {pages} {3296} (\bibinfo {year} {2008})}\BibitemShut {NoStop}%
\bibitem [{\citenamefont {Hsu}\ \emph {et~al.}(2008)\citenamefont {Hsu},
  \citenamefont {Luo}, \citenamefont {Yeh}, \citenamefont {Chen}, \citenamefont
  {Huang}, \citenamefont {Wu}, \citenamefont {Lee}, \citenamefont {Huang},
  \citenamefont {Chu}, \citenamefont {Yan},\ and\ \citenamefont
  {Wu}}]{HsuFeSe}%
  \BibitemOpen
  \bibfield  {author} {\bibinfo {author} {\bibfnamefont {F.-C.}\ \bibnamefont
  {Hsu}}, \bibinfo {author} {\bibfnamefont {J.-Y.}\ \bibnamefont {Luo}},
  \bibinfo {author} {\bibfnamefont {K.-W.}\ \bibnamefont {Yeh}}, \bibinfo
  {author} {\bibfnamefont {T.-K.}\ \bibnamefont {Chen}}, \bibinfo {author}
  {\bibfnamefont {T.-W.}\ \bibnamefont {Huang}}, \bibinfo {author}
  {\bibfnamefont {P.~M.}\ \bibnamefont {Wu}}, \bibinfo {author} {\bibfnamefont
  {Y.-C.}\ \bibnamefont {Lee}}, \bibinfo {author} {\bibfnamefont {Y.-L.}\
  \bibnamefont {Huang}}, \bibinfo {author} {\bibfnamefont {Y.-Y.}\ \bibnamefont
  {Chu}}, \bibinfo {author} {\bibfnamefont {D.-C.}\ \bibnamefont {Yan}}, \ and\
  \bibinfo {author} {\bibfnamefont {M.-K.}\ \bibnamefont {Wu}},\ }\href
  {\doibase 10.1073/pnas.0807325105} {\bibfield  {journal} {\bibinfo  {journal}
  {Proceedings of the National Academy of Sciences}\ }\textbf {\bibinfo
  {volume} {105}},\ \bibinfo {pages} {14262} (\bibinfo {year}
  {2008})}\BibitemShut {NoStop}%
\bibitem [{\citenamefont {Masaki}\ \emph {et~al.}(2009)\citenamefont {Masaki},
  \citenamefont {Kotegawa}, \citenamefont {Hara}, \citenamefont {Tou},
  \citenamefont {Murata}, \citenamefont {Mizuguchi},\ and\ \citenamefont
  {Takano}}]{JPSJ.78.063704}%
  \BibitemOpen
  \bibfield  {author} {\bibinfo {author} {\bibfnamefont {S.}~\bibnamefont
  {Masaki}}, \bibinfo {author} {\bibfnamefont {H.}~\bibnamefont {Kotegawa}},
  \bibinfo {author} {\bibfnamefont {Y.}~\bibnamefont {Hara}}, \bibinfo {author}
  {\bibfnamefont {H.}~\bibnamefont {Tou}}, \bibinfo {author} {\bibfnamefont
  {K.}~\bibnamefont {Murata}}, \bibinfo {author} {\bibfnamefont
  {Y.}~\bibnamefont {Mizuguchi}}, \ and\ \bibinfo {author} {\bibfnamefont
  {Y.}~\bibnamefont {Takano}},\ }\href {\doibase 10.1143/JPSJ.78.063704}
  {\bibfield  {journal} {\bibinfo  {journal} {Journal of the Physical Society
  of Japan}\ }\textbf {\bibinfo {volume} {78}},\ \bibinfo {pages} {063704}
  (\bibinfo {year} {2009})}\BibitemShut {NoStop}%
\bibitem [{\citenamefont {Medvedev}\ \emph {et~al.}(2009)\citenamefont
  {Medvedev}, \citenamefont {McQueen}, \citenamefont {Troyan}, \citenamefont
  {Palasyuk}, \citenamefont {Eremets}, \citenamefont {Cava}, \citenamefont
  {Naghavi}, \citenamefont {Casper}, \citenamefont {Ksenofontov}, \citenamefont
  {Wortmann},\ and\ \citenamefont {Felser}}]{nmat2491}%
  \BibitemOpen
  \bibfield  {author} {\bibinfo {author} {\bibfnamefont {S.}~\bibnamefont
  {Medvedev}}, \bibinfo {author} {\bibfnamefont {T.~M.}\ \bibnamefont
  {McQueen}}, \bibinfo {author} {\bibfnamefont {I.~A.}\ \bibnamefont {Troyan}},
  \bibinfo {author} {\bibfnamefont {T.}~\bibnamefont {Palasyuk}}, \bibinfo
  {author} {\bibfnamefont {M.~I.}\ \bibnamefont {Eremets}}, \bibinfo {author}
  {\bibfnamefont {R.~J.}\ \bibnamefont {Cava}}, \bibinfo {author}
  {\bibfnamefont {S.}~\bibnamefont {Naghavi}}, \bibinfo {author} {\bibfnamefont
  {F.}~\bibnamefont {Casper}}, \bibinfo {author} {\bibfnamefont
  {V.}~\bibnamefont {Ksenofontov}}, \bibinfo {author} {\bibfnamefont
  {G.}~\bibnamefont {Wortmann}}, \ and\ \bibinfo {author} {\bibfnamefont
  {C.}~\bibnamefont {Felser}},\ }\href {\doibase 10.1038/nmat2491} {\bibfield
  {journal} {\bibinfo  {journal} {Nature Materials}\ }\textbf {\bibinfo
  {volume} {8}},\ \bibinfo {pages} {630} (\bibinfo {year} {2009})}\BibitemShut
  {NoStop}%
\bibitem [{\citenamefont {Bellingeri}\ \emph {et~al.}(2010)\citenamefont
  {Bellingeri}, \citenamefont {Pallecchi}, \citenamefont {Buzio}, \citenamefont
  {Gerbi}, \citenamefont {Marr\`{e}}, \citenamefont {Cimberle}, \citenamefont
  {Tropeano}, \citenamefont {Putti}, \citenamefont {Palenzona},\ and\
  \citenamefont {Ferdeghini}}]{bellingeri:102512}%
  \BibitemOpen
  \bibfield  {author} {\bibinfo {author} {\bibfnamefont {E.}~\bibnamefont
  {Bellingeri}}, \bibinfo {author} {\bibfnamefont {I.}~\bibnamefont
  {Pallecchi}}, \bibinfo {author} {\bibfnamefont {R.}~\bibnamefont {Buzio}},
  \bibinfo {author} {\bibfnamefont {A.}~\bibnamefont {Gerbi}}, \bibinfo
  {author} {\bibfnamefont {D.}~\bibnamefont {Marr\`{e}}}, \bibinfo {author}
  {\bibfnamefont {M.~R.}\ \bibnamefont {Cimberle}}, \bibinfo {author}
  {\bibfnamefont {M.}~\bibnamefont {Tropeano}}, \bibinfo {author}
  {\bibfnamefont {M.}~\bibnamefont {Putti}}, \bibinfo {author} {\bibfnamefont
  {A.}~\bibnamefont {Palenzona}}, \ and\ \bibinfo {author} {\bibfnamefont
  {C.}~\bibnamefont {Ferdeghini}},\ }\href {\doibase 10.1063/1.3358148}
  {\bibfield  {journal} {\bibinfo  {journal} {Applied Physics Letters}\
  }\textbf {\bibinfo {volume} {96}},\ \bibinfo {eid} {102512} (\bibinfo {year}
  {2010})}\BibitemShut {NoStop}%
\bibitem [{\citenamefont {Iida}\ \emph {et~al.}(2011)\citenamefont {Iida},
  \citenamefont {Hanisch}, \citenamefont {Schulze}, \citenamefont {Aswartham},
  \citenamefont {Wurmehl}, \citenamefont {Buchner}, \citenamefont {Schultz},\
  and\ \citenamefont {Holzapfel}}]{iida:202503}%
  \BibitemOpen
  \bibfield  {author} {\bibinfo {author} {\bibfnamefont {K.}~\bibnamefont
  {Iida}}, \bibinfo {author} {\bibfnamefont {J.}~\bibnamefont {Hanisch}},
  \bibinfo {author} {\bibfnamefont {M.}~\bibnamefont {Schulze}}, \bibinfo
  {author} {\bibfnamefont {S.}~\bibnamefont {Aswartham}}, \bibinfo {author}
  {\bibfnamefont {S.}~\bibnamefont {Wurmehl}}, \bibinfo {author} {\bibfnamefont
  {B.}~\bibnamefont {Buchner}}, \bibinfo {author} {\bibfnamefont
  {L.}~\bibnamefont {Schultz}}, \ and\ \bibinfo {author} {\bibfnamefont
  {B.}~\bibnamefont {Holzapfel}},\ }\href {\doibase 10.1063/1.3660257}
  {\bibfield  {journal} {\bibinfo  {journal} {Applied Physics Letters}\
  }\textbf {\bibinfo {volume} {99}},\ \bibinfo {eid} {202503} (\bibinfo {year}
  {2011})}\BibitemShut {NoStop}%
\bibitem [{\citenamefont {Tsukada}\ \emph {et~al.}(2011)\citenamefont
  {Tsukada}, \citenamefont {Hanawa}, \citenamefont {Akiike}, \citenamefont
  {Nabeshima}, \citenamefont {Imai}, \citenamefont {Ichinose}, \citenamefont
  {Komiya}, \citenamefont {Hikage}, \citenamefont {Kawaguchi}, \citenamefont
  {Ikuta},\ and\ \citenamefont {Maeda}}]{APEX.4.053101}%
  \BibitemOpen
  \bibfield  {author} {\bibinfo {author} {\bibfnamefont {I.}~\bibnamefont
  {Tsukada}}, \bibinfo {author} {\bibfnamefont {M.}~\bibnamefont {Hanawa}},
  \bibinfo {author} {\bibfnamefont {T.}~\bibnamefont {Akiike}}, \bibinfo
  {author} {\bibfnamefont {F.}~\bibnamefont {Nabeshima}}, \bibinfo {author}
  {\bibfnamefont {Y.}~\bibnamefont {Imai}}, \bibinfo {author} {\bibfnamefont
  {A.}~\bibnamefont {Ichinose}}, \bibinfo {author} {\bibfnamefont
  {S.}~\bibnamefont {Komiya}}, \bibinfo {author} {\bibfnamefont
  {T.}~\bibnamefont {Hikage}}, \bibinfo {author} {\bibfnamefont
  {T.}~\bibnamefont {Kawaguchi}}, \bibinfo {author} {\bibfnamefont
  {H.}~\bibnamefont {Ikuta}}, \ and\ \bibinfo {author} {\bibfnamefont
  {A.}~\bibnamefont {Maeda}},\ }\href {\doibase 10.1143/APEX.4.053101}
  {\bibfield  {journal} {\bibinfo  {journal} {Applied Physics Express}\
  }\textbf {\bibinfo {volume} {4}},\ \bibinfo {pages} {053101} (\bibinfo {year}
  {2011})}\BibitemShut {NoStop}%
\bibitem [{\citenamefont {Si}\ \emph {et~al.}(2012)\citenamefont {Si},
  \citenamefont {Han}, \citenamefont {Shi}, \citenamefont {Ehrlich},
  \citenamefont {Jaroszynski}, \citenamefont {Goyal},\ and\ \citenamefont
  {Li}}]{WSiNatCom}%
  \BibitemOpen
  \bibfield  {author} {\bibinfo {author} {\bibfnamefont {W.}~\bibnamefont
  {Si}}, \bibinfo {author} {\bibfnamefont {S.~J.}\ \bibnamefont {Han}},
  \bibinfo {author} {\bibfnamefont {X.}~\bibnamefont {Shi}}, \bibinfo {author}
  {\bibfnamefont {S.~N.}\ \bibnamefont {Ehrlich}}, \bibinfo {author}
  {\bibfnamefont {J.}~\bibnamefont {Jaroszynski}}, \bibinfo {author}
  {\bibfnamefont {A.}~\bibnamefont {Goyal}}, \ and\ \bibinfo {author}
  {\bibfnamefont {Q.}~\bibnamefont {Li}},\ }\href {\doibase 10.1038/ncomms2337}
  {\bibfield  {journal} {\bibinfo  {journal} {Nat. Commun.}\ }\textbf {\bibinfo
  {volume} {4}},\ \bibinfo {pages} {1347} (\bibinfo {year} {2012})}\BibitemShut
  {NoStop}%
\bibitem [{\citenamefont {Horigane}\ \emph {et~al.}(2009)\citenamefont
  {Horigane}, \citenamefont {Takeshita}, \citenamefont {Lee}, \citenamefont
  {Hiraka},\ and\ \citenamefont {Yamada}}]{JPSJ.78.063705}%
  \BibitemOpen
  \bibfield  {author} {\bibinfo {author} {\bibfnamefont {K.}~\bibnamefont
  {Horigane}}, \bibinfo {author} {\bibfnamefont {N.}~\bibnamefont {Takeshita}},
  \bibinfo {author} {\bibfnamefont {C.-H.}\ \bibnamefont {Lee}}, \bibinfo
  {author} {\bibfnamefont {H.}~\bibnamefont {Hiraka}}, \ and\ \bibinfo {author}
  {\bibfnamefont {K.}~\bibnamefont {Yamada}},\ }\href {\doibase
  10.1143/JPSJ.78.063705} {\bibfield  {journal} {\bibinfo  {journal} {Journal
  of the Physical Society of Japan}\ }\textbf {\bibinfo {volume} {78}},\
  \bibinfo {pages} {063705} (\bibinfo {year} {2009})}\BibitemShut {NoStop}%
\bibitem [{\citenamefont {Qing-Yan}\ \emph {et~al.}(2012)\citenamefont
  {Qing-Yan}, \citenamefont {Zhi}, \citenamefont {Wen-Hao}, \citenamefont
  {Zuo-Cheng}, \citenamefont {Jin-Song}, \citenamefont {Wei}, \citenamefont
  {Hao}, \citenamefont {Yun-Bo}, \citenamefont {Peng}, \citenamefont {Kai},
  \citenamefont {Jing}, \citenamefont {Can-Li}, \citenamefont {Ke},
  \citenamefont {Jin-Feng}, \citenamefont {Shuai-Hua}, \citenamefont {Ya-Yu},
  \citenamefont {Li-Li}, \citenamefont {Xi}, \citenamefont {Xu-Cun},\ and\
  \citenamefont {Qi-Kun}}]{0256-307X-29-3-037402}%
  \BibitemOpen
  \bibfield  {author} {\bibinfo {author} {\bibfnamefont {W.}~\bibnamefont
  {Qing-Yan}}, \bibinfo {author} {\bibfnamefont {L.}~\bibnamefont {Zhi}},
  \bibinfo {author} {\bibfnamefont {Z.}~\bibnamefont {Wen-Hao}}, \bibinfo
  {author} {\bibfnamefont {Z.}~\bibnamefont {Zuo-Cheng}}, \bibinfo {author}
  {\bibfnamefont {Z.}~\bibnamefont {Jin-Song}}, \bibinfo {author}
  {\bibfnamefont {L.}~\bibnamefont {Wei}}, \bibinfo {author} {\bibfnamefont
  {D.}~\bibnamefont {Hao}}, \bibinfo {author} {\bibfnamefont {O.}~\bibnamefont
  {Yun-Bo}}, \bibinfo {author} {\bibfnamefont {D.}~\bibnamefont {Peng}},
  \bibinfo {author} {\bibfnamefont {C.}~\bibnamefont {Kai}}, \bibinfo {author}
  {\bibfnamefont {W.}~\bibnamefont {Jing}}, \bibinfo {author} {\bibfnamefont
  {S.}~\bibnamefont {Can-Li}}, \bibinfo {author} {\bibfnamefont
  {H.}~\bibnamefont {Ke}}, \bibinfo {author} {\bibfnamefont {J.}~\bibnamefont
  {Jin-Feng}}, \bibinfo {author} {\bibfnamefont {J.}~\bibnamefont {Shuai-Hua}},
  \bibinfo {author} {\bibfnamefont {W.}~\bibnamefont {Ya-Yu}}, \bibinfo
  {author} {\bibfnamefont {W.}~\bibnamefont {Li-Li}}, \bibinfo {author}
  {\bibfnamefont {C.}~\bibnamefont {Xi}}, \bibinfo {author} {\bibfnamefont
  {M.}~\bibnamefont {Xu-Cun}}, \ and\ \bibinfo {author} {\bibfnamefont
  {X.}~\bibnamefont {Qi-Kun}},\ }\href
  {http://stacks.iop.org/0256-307X/29/i=3/a=037402} {\bibfield  {journal}
  {\bibinfo  {journal} {Chinese Physics Letters}\ }\textbf {\bibinfo {volume}
  {29}},\ \bibinfo {pages} {037402} (\bibinfo {year} {2012})}\BibitemShut
  {NoStop}%
\bibitem [{\citenamefont {Han}\ \emph {et~al.}(2009)\citenamefont {Han},
  \citenamefont {Li}, \citenamefont {Cao}, \citenamefont {Zhang}, \citenamefont
  {Xu},\ and\ \citenamefont {Zhao}}]{0953-8984-21-23-235702}%
  \BibitemOpen
  \bibfield  {author} {\bibinfo {author} {\bibfnamefont {Y.}~\bibnamefont
  {Han}}, \bibinfo {author} {\bibfnamefont {W.~Y.}\ \bibnamefont {Li}},
  \bibinfo {author} {\bibfnamefont {L.~X.}\ \bibnamefont {Cao}}, \bibinfo
  {author} {\bibfnamefont {S.}~\bibnamefont {Zhang}}, \bibinfo {author}
  {\bibfnamefont {B.}~\bibnamefont {Xu}}, \ and\ \bibinfo {author}
  {\bibfnamefont {B.~R.}\ \bibnamefont {Zhao}},\ }\href
  {http://stacks.iop.org/0953-8984/21/i=23/a=235702} {\bibfield  {journal}
  {\bibinfo  {journal} {Journal of Physics: Condensed Matter}\ }\textbf
  {\bibinfo {volume} {21}},\ \bibinfo {pages} {235702} (\bibinfo {year}
  {2009})}\BibitemShut {NoStop}%
\bibitem [{\citenamefont {Nie}\ \emph {et~al.}(2009)\citenamefont {Nie},
  \citenamefont {Brahimi}, \citenamefont {Budnick}, \citenamefont {Hines},
  \citenamefont {Jain},\ and\ \citenamefont {Wells}}]{nie:242505}%
  \BibitemOpen
  \bibfield  {author} {\bibinfo {author} {\bibfnamefont {Y.~F.}\ \bibnamefont
  {Nie}}, \bibinfo {author} {\bibfnamefont {E.}~\bibnamefont {Brahimi}},
  \bibinfo {author} {\bibfnamefont {J.~I.}\ \bibnamefont {Budnick}}, \bibinfo
  {author} {\bibfnamefont {W.~A.}\ \bibnamefont {Hines}}, \bibinfo {author}
  {\bibfnamefont {M.}~\bibnamefont {Jain}}, \ and\ \bibinfo {author}
  {\bibfnamefont {B.~O.}\ \bibnamefont {Wells}},\ }\href {\doibase
  10.1063/1.3155441} {\bibfield  {journal} {\bibinfo  {journal} {Applied
  Physics Letters}\ }\textbf {\bibinfo {volume} {94}},\ \bibinfo {eid} {242505}
  (\bibinfo {year} {2009})}\BibitemShut {NoStop}%
\bibitem [{\citenamefont {Wang}\ \emph {et~al.}(2009)\citenamefont {Wang},
  \citenamefont {Luo}, \citenamefont {Huang}, \citenamefont {Chang},
  \citenamefont {Chen}, \citenamefont {Hsu}, \citenamefont {Wu}, \citenamefont
  {Wu}, \citenamefont {Chang},\ and\ \citenamefont
  {Wu}}]{PhysRevLett.103.117002}%
  \BibitemOpen
  \bibfield  {author} {\bibinfo {author} {\bibfnamefont {M.~J.}\ \bibnamefont
  {Wang}}, \bibinfo {author} {\bibfnamefont {J.~Y.}\ \bibnamefont {Luo}},
  \bibinfo {author} {\bibfnamefont {T.~W.}\ \bibnamefont {Huang}}, \bibinfo
  {author} {\bibfnamefont {H.~H.}\ \bibnamefont {Chang}}, \bibinfo {author}
  {\bibfnamefont {T.~K.}\ \bibnamefont {Chen}}, \bibinfo {author}
  {\bibfnamefont {F.~C.}\ \bibnamefont {Hsu}}, \bibinfo {author} {\bibfnamefont
  {C.~T.}\ \bibnamefont {Wu}}, \bibinfo {author} {\bibfnamefont {P.~M.}\
  \bibnamefont {Wu}}, \bibinfo {author} {\bibfnamefont {A.~M.}\ \bibnamefont
  {Chang}}, \ and\ \bibinfo {author} {\bibfnamefont {M.~K.}\ \bibnamefont
  {Wu}},\ }\href {\doibase 10.1103/PhysRevLett.103.117002} {\bibfield
  {journal} {\bibinfo  {journal} {Phys. Rev. Lett.}\ }\textbf {\bibinfo
  {volume} {103}},\ \bibinfo {pages} {117002} (\bibinfo {year}
  {2009})}\BibitemShut {NoStop}%
\bibitem [{\citenamefont {Jourdan}\ and\ \citenamefont {ten
  Haaf}(2010)}]{jourdan:023913}%
  \BibitemOpen
  \bibfield  {author} {\bibinfo {author} {\bibfnamefont {M.}~\bibnamefont
  {Jourdan}}\ and\ \bibinfo {author} {\bibfnamefont {S.}~\bibnamefont {ten
  Haaf}},\ }\href {\doibase 10.1063/1.3465082} {\bibfield  {journal} {\bibinfo
  {journal} {Journal of Applied Physics}\ }\textbf {\bibinfo {volume} {108}},\
  \bibinfo {eid} {023913} (\bibinfo {year} {2010})}\BibitemShut {NoStop}%
\bibitem [{\citenamefont {Jung}\ \emph {et~al.}(2010)\citenamefont {Jung},
  \citenamefont {Lee}, \citenamefont {Choi}, \citenamefont {Kang},
  \citenamefont {Lee}, \citenamefont {Hwang},\ and\ \citenamefont
  {Kim}}]{Jung20101977}%
  \BibitemOpen
  \bibfield  {author} {\bibinfo {author} {\bibfnamefont {S.-G.}\ \bibnamefont
  {Jung}}, \bibinfo {author} {\bibfnamefont {N.}~\bibnamefont {Lee}}, \bibinfo
  {author} {\bibfnamefont {E.-M.}\ \bibnamefont {Choi}}, \bibinfo {author}
  {\bibfnamefont {W.}~\bibnamefont {Kang}}, \bibinfo {author} {\bibfnamefont
  {S.-I.}\ \bibnamefont {Lee}}, \bibinfo {author} {\bibfnamefont {T.-J.}\
  \bibnamefont {Hwang}}, \ and\ \bibinfo {author} {\bibfnamefont
  {D.}~\bibnamefont {Kim}},\ }\href {\doibase 10.1016/j.physc.2010.08.011}
  {\bibfield  {journal} {\bibinfo  {journal} {Physica C: Superconductivity}\
  }\textbf {\bibinfo {volume} {470}},\ \bibinfo {pages} {1977 } (\bibinfo
  {year} {2010})}\BibitemShut {NoStop}%
\bibitem [{\citenamefont {Chen}\ \emph {et~al.}(2011)\citenamefont {Chen},
  \citenamefont {Tsai}, \citenamefont {Zhu}, \citenamefont {Bi},\ and\
  \citenamefont {Wang}}]{Chen2011515}%
  \BibitemOpen
  \bibfield  {author} {\bibinfo {author} {\bibfnamefont {L.}~\bibnamefont
  {Chen}}, \bibinfo {author} {\bibfnamefont {C.-F.}\ \bibnamefont {Tsai}},
  \bibinfo {author} {\bibfnamefont {Y.}~\bibnamefont {Zhu}}, \bibinfo {author}
  {\bibfnamefont {Z.}~\bibnamefont {Bi}}, \ and\ \bibinfo {author}
  {\bibfnamefont {H.}~\bibnamefont {Wang}},\ }\href {\doibase
  10.1016/j.physc.2011.05.248} {\bibfield  {journal} {\bibinfo  {journal}
  {Physica C: Superconductivity}\ }\textbf {\bibinfo {volume} {471}},\ \bibinfo
  {pages} {515 } (\bibinfo {year} {2011})}\BibitemShut {NoStop}%
\bibitem [{\citenamefont {Schneider}\ \emph {et~al.}(2012)\citenamefont
  {Schneider}, \citenamefont {Zaitsev}, \citenamefont {Fuchs},\ and\
  \citenamefont {v.~L\"ohneysen}}]{PhysRevLett.108.257003}%
  \BibitemOpen
  \bibfield  {author} {\bibinfo {author} {\bibfnamefont {R.}~\bibnamefont
  {Schneider}}, \bibinfo {author} {\bibfnamefont {A.~G.}\ \bibnamefont
  {Zaitsev}}, \bibinfo {author} {\bibfnamefont {D.}~\bibnamefont {Fuchs}}, \
  and\ \bibinfo {author} {\bibfnamefont {H.}~\bibnamefont {v.~L\"ohneysen}},\
  }\href {\doibase 10.1103/PhysRevLett.108.257003} {\bibfield  {journal}
  {\bibinfo  {journal} {Phys. Rev. Lett.}\ }\textbf {\bibinfo {volume} {108}},\
  \bibinfo {pages} {257003} (\bibinfo {year} {2012})}\BibitemShut {NoStop}%
\bibitem [{\citenamefont {Imai}\ \emph
  {et~al.}(2010{\natexlab{a}})\citenamefont {Imai}, \citenamefont {Akiike},
  \citenamefont {Hanawa}, \citenamefont {Tsukada}, \citenamefont {Ichinose},
  \citenamefont {Maeda}, \citenamefont {Hikage}, \citenamefont {Kawaguchi},\
  and\ \citenamefont {Ikuta}}]{APEX.3.043102}%
  \BibitemOpen
  \bibfield  {author} {\bibinfo {author} {\bibfnamefont {Y.}~\bibnamefont
  {Imai}}, \bibinfo {author} {\bibfnamefont {T.}~\bibnamefont {Akiike}},
  \bibinfo {author} {\bibfnamefont {M.}~\bibnamefont {Hanawa}}, \bibinfo
  {author} {\bibfnamefont {I.}~\bibnamefont {Tsukada}}, \bibinfo {author}
  {\bibfnamefont {A.}~\bibnamefont {Ichinose}}, \bibinfo {author}
  {\bibfnamefont {A.}~\bibnamefont {Maeda}}, \bibinfo {author} {\bibfnamefont
  {T.}~\bibnamefont {Hikage}}, \bibinfo {author} {\bibfnamefont
  {T.}~\bibnamefont {Kawaguchi}}, \ and\ \bibinfo {author} {\bibfnamefont
  {H.}~\bibnamefont {Ikuta}},\ }\href {\doibase 10.1143/APEX.3.043102}
  {\bibfield  {journal} {\bibinfo  {journal} {Applied Physics Express}\
  }\textbf {\bibinfo {volume} {3}},\ \bibinfo {pages} {043102} (\bibinfo {year}
  {2010}{\natexlab{a}})}\BibitemShut {NoStop}%
\bibitem [{\citenamefont {Imai}\ \emph
  {et~al.}(2010{\natexlab{b}})\citenamefont {Imai}, \citenamefont {Tanaka},
  \citenamefont {Akiike}, \citenamefont {Hanawa}, \citenamefont {Tsukada},\
  and\ \citenamefont {Maeda}}]{JJAP.49.023101}%
  \BibitemOpen
  \bibfield  {author} {\bibinfo {author} {\bibfnamefont {Y.}~\bibnamefont
  {Imai}}, \bibinfo {author} {\bibfnamefont {R.}~\bibnamefont {Tanaka}},
  \bibinfo {author} {\bibfnamefont {T.}~\bibnamefont {Akiike}}, \bibinfo
  {author} {\bibfnamefont {M.}~\bibnamefont {Hanawa}}, \bibinfo {author}
  {\bibfnamefont {I.}~\bibnamefont {Tsukada}}, \ and\ \bibinfo {author}
  {\bibfnamefont {A.}~\bibnamefont {Maeda}},\ }\href {\doibase
  10.1143/JJAP.49.023101} {\bibfield  {journal} {\bibinfo  {journal} {Japanese
  Journal of Applied Physics}\ }\textbf {\bibinfo {volume} {49}},\ \bibinfo
  {pages} {023101} (\bibinfo {year} {2010}{\natexlab{b}})}\BibitemShut
  {NoStop}%
\bibitem [{\citenamefont {McQueen}\ \emph {et~al.}(2009)\citenamefont
  {McQueen}, \citenamefont {Huang}, \citenamefont {Ksenofontov}, \citenamefont
  {Felser}, \citenamefont {Xu}, \citenamefont {Zandbergen}, \citenamefont
  {Hor}, \citenamefont {Allred}, \citenamefont {Williams}, \citenamefont {Qu},
  \citenamefont {Checkelsky}, \citenamefont {Ong},\ and\ \citenamefont
  {Cava}}]{PhysRevB.79.014522}%
  \BibitemOpen
  \bibfield  {author} {\bibinfo {author} {\bibfnamefont {T.~M.}\ \bibnamefont
  {McQueen}}, \bibinfo {author} {\bibfnamefont {Q.}~\bibnamefont {Huang}},
  \bibinfo {author} {\bibfnamefont {V.}~\bibnamefont {Ksenofontov}}, \bibinfo
  {author} {\bibfnamefont {C.}~\bibnamefont {Felser}}, \bibinfo {author}
  {\bibfnamefont {Q.}~\bibnamefont {Xu}}, \bibinfo {author} {\bibfnamefont
  {H.}~\bibnamefont {Zandbergen}}, \bibinfo {author} {\bibfnamefont {Y.~S.}\
  \bibnamefont {Hor}}, \bibinfo {author} {\bibfnamefont {J.}~\bibnamefont
  {Allred}}, \bibinfo {author} {\bibfnamefont {A.~J.}\ \bibnamefont
  {Williams}}, \bibinfo {author} {\bibfnamefont {D.}~\bibnamefont {Qu}},
  \bibinfo {author} {\bibfnamefont {J.}~\bibnamefont {Checkelsky}}, \bibinfo
  {author} {\bibfnamefont {N.~P.}\ \bibnamefont {Ong}}, \ and\ \bibinfo
  {author} {\bibfnamefont {R.~J.}\ \bibnamefont {Cava}},\ }\href {\doibase
  10.1103/PhysRevB.79.014522} {\bibfield  {journal} {\bibinfo  {journal} {Phys.
  Rev. B}\ }\textbf {\bibinfo {volume} {79}},\ \bibinfo {pages} {014522}
  (\bibinfo {year} {2009})}\BibitemShut {NoStop}%
\bibitem [{\citenamefont {de~Souza}\ \emph {et~al.}(2010)\citenamefont
  {de~Souza}, \citenamefont {Haghighirad}, \citenamefont {Tutsch},
  \citenamefont {Assmus},\ and\ \citenamefont {Lang}}]{Souza}%
  \BibitemOpen
  \bibfield  {author} {\bibinfo {author} {\bibfnamefont {M.}~\bibnamefont
  {de~Souza}}, \bibinfo {author} {\bibfnamefont {A.-A.}\ \bibnamefont
  {Haghighirad}}, \bibinfo {author} {\bibfnamefont {U.}~\bibnamefont {Tutsch}},
  \bibinfo {author} {\bibfnamefont {W.}~\bibnamefont {Assmus}}, \ and\ \bibinfo
  {author} {\bibfnamefont {M.}~\bibnamefont {Lang}},\ }\href {\doibase
  10.1140/epjb/e2010-00254-7} {\bibfield  {journal} {\bibinfo  {journal} {The
  European Physical Journal B}\ }\textbf {\bibinfo {volume} {77}},\ \bibinfo
  {pages} {101} (\bibinfo {year} {2010})}\BibitemShut {NoStop}%
\bibitem [{\citenamefont {Hu}\ \emph {et~al.}(2011)\citenamefont {Hu},
  \citenamefont {Lei}, \citenamefont {Abeykoon}, \citenamefont {Bozin},
  \citenamefont {Billinge}, \citenamefont {Warren}, \citenamefont {Siegrist},\
  and\ \citenamefont {Petrovic}}]{PhysRevB.83.224502}%
  \BibitemOpen
  \bibfield  {author} {\bibinfo {author} {\bibfnamefont {R.}~\bibnamefont
  {Hu}}, \bibinfo {author} {\bibfnamefont {H.}~\bibnamefont {Lei}}, \bibinfo
  {author} {\bibfnamefont {M.}~\bibnamefont {Abeykoon}}, \bibinfo {author}
  {\bibfnamefont {E.~S.}\ \bibnamefont {Bozin}}, \bibinfo {author}
  {\bibfnamefont {S.~J.~L.}\ \bibnamefont {Billinge}}, \bibinfo {author}
  {\bibfnamefont {J.~B.}\ \bibnamefont {Warren}}, \bibinfo {author}
  {\bibfnamefont {T.}~\bibnamefont {Siegrist}}, \ and\ \bibinfo {author}
  {\bibfnamefont {C.}~\bibnamefont {Petrovic}},\ }\href {\doibase
  10.1103/PhysRevB.83.224502} {\bibfield  {journal} {\bibinfo  {journal} {Phys.
  Rev. B}\ }\textbf {\bibinfo {volume} {83}},\ \bibinfo {pages} {224502}
  (\bibinfo {year} {2011})}\BibitemShut {NoStop}%
\bibitem [{\citenamefont {Ichinose}\ \emph {et~al.}(2013)\citenamefont
  {Ichinose}, \citenamefont {Nabeshima}, \citenamefont {Tsukada}, \citenamefont
  {Hanawa}, \citenamefont {Komiya}, \citenamefont {Akiike}, \citenamefont
  {Imai},\ and\ \citenamefont {Maeda}}]{IchinoseCaF2}%
  \BibitemOpen
  \bibfield  {author} {\bibinfo {author} {\bibfnamefont {A.}~\bibnamefont
  {Ichinose}}, \bibinfo {author} {\bibfnamefont {F.}~\bibnamefont {Nabeshima}},
  \bibinfo {author} {\bibfnamefont {I.}~\bibnamefont {Tsukada}}, \bibinfo
  {author} {\bibfnamefont {M.}~\bibnamefont {Hanawa}}, \bibinfo {author}
  {\bibfnamefont {S.}~\bibnamefont {Komiya}}, \bibinfo {author} {\bibfnamefont
  {T.}~\bibnamefont {Akiike}}, \bibinfo {author} {\bibfnamefont
  {Y.}~\bibnamefont {Imai}}, \ and\ \bibinfo {author} {\bibfnamefont
  {A.}~\bibnamefont {Maeda}},\ }\href
  {http://stacks.iop.org/0953-2048/26/i=7/a=075002} {\bibfield  {journal}
  {\bibinfo  {journal} {Superconductor Science and Technology}\ }\textbf
  {\bibinfo {volume} {26}},\ \bibinfo {pages} {075002} (\bibinfo {year}
  {2013})}\BibitemShut {NoStop}%
\bibitem [{\citenamefont {Imai}\ \emph {et~al.}(2012)\citenamefont {Imai},
  \citenamefont {Nabeshima}, \citenamefont {Yoshinaka}, \citenamefont
  {Miyatani}, \citenamefont {Kondo}, \citenamefont {Komiya}, \citenamefont
  {Tsukada},\ and\ \citenamefont {Maeda}}]{Bi2Pd}%
  \BibitemOpen
  \bibfield  {author} {\bibinfo {author} {\bibfnamefont {Y.}~\bibnamefont
  {Imai}}, \bibinfo {author} {\bibfnamefont {F.}~\bibnamefont {Nabeshima}},
  \bibinfo {author} {\bibfnamefont {T.}~\bibnamefont {Yoshinaka}}, \bibinfo
  {author} {\bibfnamefont {K.}~\bibnamefont {Miyatani}}, \bibinfo {author}
  {\bibfnamefont {R.}~\bibnamefont {Kondo}}, \bibinfo {author} {\bibfnamefont
  {S.}~\bibnamefont {Komiya}}, \bibinfo {author} {\bibfnamefont
  {I.}~\bibnamefont {Tsukada}}, \ and\ \bibinfo {author} {\bibfnamefont
  {A.}~\bibnamefont {Maeda}},\ }\href {\doibase 10.1143/JPSJ.81.113708}
  {\bibfield  {journal} {\bibinfo  {journal} {Journal of the Physical Society
  of Japan}\ }\textbf {\bibinfo {volume} {81}},\ \bibinfo {pages} {113708}
  (\bibinfo {year} {2012})}\BibitemShut {NoStop}%
\bibitem [{\citenamefont {Lee}\ \emph {et~al.}(2008)\citenamefont {Lee},
  \citenamefont {Iyo}, \citenamefont {Eisaki}, \citenamefont {Kito},
  \citenamefont {Fernandez-Diaz}, \citenamefont {Ito}, \citenamefont {Kihou},
  \citenamefont {Matsuhata}, \citenamefont {Braden},\ and\ \citenamefont
  {Yamada}}]{JPSJ.77.083704}%
  \BibitemOpen
  \bibfield  {author} {\bibinfo {author} {\bibfnamefont {C.-H.}\ \bibnamefont
  {Lee}}, \bibinfo {author} {\bibfnamefont {A.}~\bibnamefont {Iyo}}, \bibinfo
  {author} {\bibfnamefont {H.}~\bibnamefont {Eisaki}}, \bibinfo {author}
  {\bibfnamefont {H.}~\bibnamefont {Kito}}, \bibinfo {author} {\bibfnamefont
  {M.~T.}\ \bibnamefont {Fernandez-Diaz}}, \bibinfo {author} {\bibfnamefont
  {T.}~\bibnamefont {Ito}}, \bibinfo {author} {\bibfnamefont {K.}~\bibnamefont
  {Kihou}}, \bibinfo {author} {\bibfnamefont {H.}~\bibnamefont {Matsuhata}},
  \bibinfo {author} {\bibfnamefont {M.}~\bibnamefont {Braden}}, \ and\ \bibinfo
  {author} {\bibfnamefont {K.}~\bibnamefont {Yamada}},\ }\href {\doibase
  10.1143/JPSJ.77.083704} {\bibfield  {journal} {\bibinfo  {journal} {Journal
  of the Physical Society of Japan}\ }\textbf {\bibinfo {volume} {77}},\
  \bibinfo {pages} {083704} (\bibinfo {year} {2008})}\BibitemShut {NoStop}%
\bibitem [{\citenamefont {Mizuguchi}\ and\ \citenamefont
  {Takano}(2010)}]{JPSJ.79.102001}%
  \BibitemOpen
  \bibfield  {author} {\bibinfo {author} {\bibfnamefont {Y.}~\bibnamefont
  {Mizuguchi}}\ and\ \bibinfo {author} {\bibfnamefont {Y.}~\bibnamefont
  {Takano}},\ }\href {\doibase 10.1143/JPSJ.79.102001} {\bibfield  {journal}
  {\bibinfo  {journal} {Journal of the Physical Society of Japan}\ }\textbf
  {\bibinfo {volume} {79}},\ \bibinfo {pages} {102001} (\bibinfo {year}
  {2010})}\BibitemShut {NoStop}%
\end{thebibliography}
\end{document}